# Quantitative Determination of the Pairing Interactions for High Temperature Superconductivity in Cuprates

Jin Mo Bok [1,2], Jong Ju Bae [1], Han-Yong Choi [1,3,*], Chandra M. Varma [4*], Wentao Zhang [2,5],
Junfeng He [2], Yuxiao Zhang [2], Li Yu [2]  and X. J. Zhou [2,6*]

[1]*Department of Physics and Institute for Basic Science Research,*
*SungKyunKwan University,*
*Suwon 440-746, Korea.*
[2]*National Lab for Superconductivity,*
*Beijing National Laboratory for Condensed Matter Physics,*
*Institute of Physics, Chinese Academy of Sciences, Beijing 100190, China*
[3]*Asia Pacific Center for Theoretical Physics, Pohang 790-784, Korea.*
[4]*Department of Physics and Astronomy,*
*University of California, Riverside, California 92521.*
[5]*Department of Physics and Astronomy,*
*Shanghai JiaoTong University, Shanghai 200240, China.*
[6]*Collaborative Innovation Center of Quantum Matter, Beijing 100871, China*
*[*]Corresponding author: hychoi@skku.ac.kr,*
*chandra.varma@ucr.edu and XJZhou@aphy.iphy.ac.cn*

**Abtract**

A profound problem in modern condensed matter physics is discovering and understanding the nature of the fluctuations and their coupling to fermions in cuprates which lead to high temperature superconductivity and the invariably associated strange metal state. Here we report the quantitative determination of the normal and pairing self-energies, made possible by laser-based angle-resolved photoemission measurements with unprecedented accuracy and stability. Through a precise inversion procedure, both the effective interactions in the attractive d-wave symmetry and the repulsive part in the full symmetry are determined. The latter are nearly angle independent. Near $T_c$ both interactions are *nearly independent of frequency, and have almost the same magnitude*, over the complete energy range of up to about 0.4 eV except for a low energy feature around 50 meV present only in the repulsive part which has less than 10% of the total spectral weight. Well below $T_c$, they both change similarly by superconductivity induced features at low energies. Besides finding the pairing self-energy and the attractive interactions for the first time, these results expose a central paradox of the high $T_c$ problem: how the same frequency independent fluctuations can dominantly scatter at angles $\pm\pi/2$ in the attractive channel as well as lead to angle-independent repulsive scattering. The experimental results are compared with the available theoretical calculations based on antiferromagnetic fluctuations, Hubbard model and the quantum-critical fluctuations of loop-current order.

Historically, the quantitative analysis by McMillan and Rowell [1] of very precise tunneling experiments, using the Eliashberg theory [2,3], decisively confirmed that the exchange of phonons by the fermions is responsible for the conventional s-wave pairing in metals such as Pb. Tunneling experiments integrate over the momentum dependence of the many body effects. This is sufficient for s-wave superconductors because the normal and superconducting properties have the full symmetry of the lattice. For any superconductor, the dependence on the momentum $\mathbf{k}$ of the normal self-energy $\Sigma(\mathbf{k}, \omega)$ has the full symmetry of the lattice but for high temperature superconductors such as the cuprates, the dependence on $\mathbf{k}$ of the pairing self-energy $\phi(\mathbf{k}, \omega)$ has a $B_{1g}$ or $d_{x^2-y^2}$ symmetry. Correspondingly, the effective interaction spectrum is characterized by two functions, (i) with the full symmetry of the lattice which we will call the normal Eliashberg function $\mathcal{E}_N(\mathbf{k}, \omega)$, and (ii) with the pairing symmetry which we will call the pairing Eliashberg function $\mathcal{E}_P(\mathbf{k}, \omega)$, rather than the single function sufficient for s-wave pairing, often denoted by $\alpha^2 F(\omega)$ [1]. The much more sophisticated angle-resolved photoemission spectroscopy (ARPES) experiments [4,5] are then required, because both the momentum dependence and the frequency dependence of the interactions are necessary to decipher the fundamental physics. It is also important to show that the procedure to determine the fluctuations using the Eliashberg theory remains valid for the pairing mediated by collective fluctuations even when their high energy cut-off is comparable to the electronic band-width.

**Results of ARPES Experiments**

The procedure to extract the normal and pairing self-energies is described in Supplementary Section **SI**. (For brevity we will omit the phrase "Supplementary Section" from here on.) The requirements for ARPES to yield the electron self-energies quantitatively are very demanding. We seek to measure the absolute magnitude of the photo-electron current per unit-photon flux at various temperatures above and below $T_c$, at various angles, and across a range of frequency extending to the upper cut-off of the fluctuations. The requirements on the data, the estimate of both the signal to noise errors and the systematic errors, how to partially correct for the latter and the limits of validity of our results and analysis are given in **SII**.

We have carried out high resolution laser-ARPES measurements on two Bi$_2$Sr$_2$CaCu$_2$O$_{8+\delta}$ (Bi2212) samples, one slightly underdoped with a $T_c$ at 89 K (UD89 hereafter) and the other overdoped with a $T_c$ at 82 K (OD82 hereafter), along various momentum cuts and at various temperatures. Some of the data on UD89 [6] and OD82 [7], (only for energy $\lesssim 0.1$ eV below $T_c$) were reported earlier but without the analysis (and extension to higher energy) that is essential for deciphering the physics, presented in this work. Figure 1 shows an example of the measured data of the photoemission intensity as a function of momentum and energy, along a momentum cut marked in the inset of Fig. 1B, at temperatures well below $T_c$ (Fig. 1A) and above $T_c$ (Fig. 1B) in the UD89 sample. As shown in Fig. 1C, the data at different temperatures overlap each other at high binding energies extremely well, showing its high-quality and reproducibility. For some momentum cuts where the dispersions show small drifts with temperature and for other systematic errors, we have employed a method of data correction (see **SII**).

The strategy suggested [8] to extract the many-body effects relies on the momentum distribution curves (MDCs), which represent the intensity of photo-electrons as a function of momentum $k_{\perp}$ perpendicular to the Fermi-surface for various fixed energies, for example, across the horizontal cuts in Fig. 1A or Fig. 1B. The two-dimensional momentum **k** is represented by the angle $\theta$ with respect to the crystalline axis and the magnitude $k_{\perp}$ measure from the $(\pi/a, \pi/a)$ point as shown in Fig. 2C . In Fig. 2A we present the measured MDC for UD89 for one of the trajectories across the Fermi-surface at five energies $\omega$ in the normal state above $T_c$ (red circles), and in the superconducting state at 16 K (blue circles). These MDCs are from more than 5,000 such plots taken; the results presented here come from the analysis of such plots in the two samples at various temperatures, angles, and energies. The signal to noise of the fit in Fig. 2A may best be appreciated from Fig. 2B where the normalized difference of the measured MDC intensities between 16 K and 97 K are compared to the same function calculated from the fits in Fig. 2A. This represents the best results we have obtained; acceptable results are shown for the OD82 sample in Fig. S1 of **SI**.

**Normal and Pairing Self-Energies**

The relation of the measured photo-electron intensity $I(\mathbf{k}, \omega)$ to the spectral function $A(\mathbf{k}, \omega)$ is described in **SI**, where we also present the procedure to extract the normal self-energy $\Sigma(\mathbf{k}, \omega)$ and the pairing self-energy $\phi(\mathbf{k}, \omega)$ by fitting the MDCs. Representative fits are shown in Fig. 2A. The MDCs in the normal state in Fig. 2A (red circles) over a wide region of energy and momentum are very well represented by Lorentzians as a function of $k_\perp$. This is true precisely[9] only if $\Sigma(\mathbf{k}, \omega)$ is a function only of $\theta$ and $\omega$. In the superconducting state, we fit the MDCs with $\phi(\mathbf{k}, \omega)$ depending only on $\theta$ and $\omega$, with almost equally good results, as shown in Fig. 2A (blue circles). As a further measure of the confidence in the data and the determination of the self-energies, we compare the measured photoemission spectrum (energy distribution curve, EDC) at a fixed momentum to the EDC calculated from our fit to the MDCs at various energies in Fig. 2D.

We present the self-energies obtained directly from such fits in Fig. 3 so that the signal to noise ratio and the limits on consistency of the data are directly visible. The evolution of the magnitude of the extracted normal self-energy and pairing self-energy is shown as a function of energy at various temperatures in Fig. 3A and 3B, respectively, for $\theta = 20°$ in OD82. The pairing self-energy measured at various $\theta$ at a temperature of 16 K for the UD89 sample is shown in panel Fig. 3C. Note that $\phi(\theta, \omega)$ has been scaled by $\cos(2\theta)$. Within the uncertainties in the data, the conclusions are the same if we scale instead by the appropriate d-wave basis for a square-lattice, $\left[\cos(k_x a/\pi) - \cos(k_y a/\pi)\right]$.

The self-energies near $T_c$ are sufficient to deduce the effective interactions leading to the value of $T_c$. Near $T_c$, the real and imaginary parts of both the normal and pairing self-energies are smooth functions of energies up to high energies (and angles). But as $T$ decreases below $T_c$, one finds in Fig. 3 two low-energy features below $\sim 75$ meV. It was suggested [10] and experimentally shown [11] that forward scattering [9] from impurities lying in between the $CuO_2$ planes produces the low energy peak at $\sim 15$ meV in the self-energies in the superconducting state. The other structure at about 65 meV is expected, for all cases in which the fluctuations are due to the interactions among the fermions themselves[12, 13]. This is because the opening of the superconducting gap $\Delta$ diminishes the spectral weight at energies below $O(2\Delta)$ and piles it up at higher energies. This process occurs in addition to the generalization to d-wave superconductors of the shift of the self-energies by $\Delta$ that is well-known in phonon mediated s-wave superconductors. These two superconductivity-induced features are irrelevant in determining the value of $T_c$, although quite important in determining the temperature dependence of the superconducting gap, which is not our focus in this paper.

In Fig. 3A, except for the low energy features, the normal (real and imaginary) self-energy

$\Sigma(\theta, \omega)$ are nearly independent of temperature. The imaginary part $\Sigma_2(\theta, \omega)$ varies linearly with $\omega$ to a good approximation, as in the normal state. In the same energy range, $\Sigma_2(\theta, \omega)$ is also nearly independent of $\theta$ as already noted in Refs. (14-16). On the other hand, not surprisingly, the real and imaginary parts of the pairing self-energy $\phi(\theta, \omega)$ in Fig. 3B systematically increases with decreasing temperature below $T_c$ with saturation at low temperatures. Except for the low energy features, the imaginary part of $\phi(\theta, \omega)$ in Fig. 3B is weakly $\omega$ dependent up to about 0.2 eV, beyond which signal to noise level does not allow quantitative conclusions. In Fig. 3C, we show that the real and imaginary parts of the pairing self energy $\phi(\theta, \omega)$ scaled by $\cos(2\theta)$ over the energy range up to $\sim 0.2$ eV are also independent of $\theta$ to about $\pm 10\%$ in an absolute value. This is an important check on the data and analysis, since in these experiments, this is the quantity deduced with the largest error because it comes from the difference between the spectra below and above $T_c$. At $\theta = 20°$, the bottom of the band from the Fermi-energy is $\sim 0.2$ eV [17], which serves as the natural cut-off. Therefore, the complete fluctuation spectra have been accessed for this angle. From the measurements of $\Sigma(\theta, \omega)$ above $T_c$ in Ref. 16 and below $T_c$ found here, the cut-off energy increases smoothly with increasing $\theta$ ($to \sim 0.4$ eV for $\theta = 45°$). It appears reasonable to assume that the pairing self-energy at other angles has the same cut-off as that of the normal self-energy, as it does at $\theta = 20°$; this can be verified by future experiments with better signal to noise for $\phi(\theta, \omega)$ at higher energies and $\theta$ closer to 45°, and $T$ closer to $T_c$.

**Normal and Pairing Eliashberg Functions**

Important conclusions about the fundamental physics of the cuprates can already be reached from the self-energies (Fig. 3), which have been directly extracted from the experimental data. Reaching some other conclusions requires solving the Eliashberg integral equations for anisotropic superconductivity, described in Supplementary SIII. We show in **SIII** that, using the experimentally obtained normal and pairing self-energies as inputs, the determination of the Eliashberg functions $\mathcal{E}_N(\theta, \omega)$ and $\mathcal{E}_P(\theta, \omega)$ is limited only by the accuracy of the self-energies and the procedure of solving the integral equations. We will also provide a self-consistency check below on the validity of this procedure using the experimental results.

$\mathcal{E}_N(\theta, \omega)$ and $\mathcal{E}_P(\theta, \omega)$ are deduced from the measured self-energies $\Sigma(\theta, \omega)$ and $\phi(\theta, \omega)$, respectively, in Fig. 3 through solution of the integral equations by the maximum entropy method [18]. In order to avoid instabilities in the numerical solutions of these equations by such a procedure, the raw data of self-energies, such as those shown in Fig. 3A-C, are smoothened at each energy by averaging over $\pm 5$ meV around it as exemplified in Fig. 3D. The results obtained are shown in Fig. 4, for both UD89 and OD82. We note that, despite the smoothening of the self-energy, weak oscillations with magnitudes of about 10% in $\mathcal{E}_N$ and $\mathcal{E}_P$ are found which are artifacts from the maximum entropy method. In Fig. 4, we plot $\mathcal{E}_N(\theta, \omega)$ and the scaled quantity $\tilde{\mathcal{E}}_P(\theta, \omega) \equiv \mathcal{E}_P(\theta, \omega)/\cos(2\theta)$.

Let us start with Fig. 4C which gives results close to $T_c$. The normal state bump at $\sim 50$ meV in $E_N(\theta, \omega)$ hardly changes for $T \lesssim T_c$ and is *absent* in $\tilde{\mathcal{E}}_P(\theta, \omega)$. Even more importantly, if we ignore the bump, $\tilde{\mathcal{E}}_P(\theta, \omega) \approx \mathcal{E}_N(\theta, \omega)$ to within 10% accuracy. As $T$ decreases well below $T_c$, both functions develop a peak in the region around 50-75 meV as shown in Fig. 4A-B. These are related to the superconductivity induced features in the self-energies that we have already discussed. The result that $\tilde{\mathcal{E}}_P(\theta, \omega)$ loses the low energy feature as $T \rightarrow T_c$ is highlighted in panel Fig. 4F, where its evolution with temperature is shown.

For comparison, we also include $\mathcal{E}_N(\theta, \omega)$ deduced from its self-energy in the normal state at various angles by the same methods in Fig. 4D-E. These results are consistent with the earlier deductions [19] from ARPES data along the diagonal ($\theta = 45°$) direction by the same technique as well as by fitting the measured optical spectra in the normal state [20, 21], which are averages over all angles weighted by their Fermi-velocity.

In Fig. 5, we have calculated the real and imaginary parts of the pairing self-energy $\phi(\theta = 20°, \omega)$ from the deduced $\mathcal{E}_N$ at 70 K (Fig. 4C) and at 35 K (Fig. 4B) by using the Eliashberg equations and assuming $\tilde{\mathcal{E}}_P = \mathcal{E}_N$. The calculations (solid lines in Fig. 5) are directly compared with the extracted values (circles and squares in Fig. 5). Because $\phi \ll \Sigma$ near $T_c$, $\mathcal{E}_N$ here is determined primarily by $\Sigma$. The measured $\phi$ determines the deduced $\mathcal{E}_P$ by the Eliashberg equations. Therefore, the success of the comparison depends both on (i) $\mathcal{E}_N \approx \tilde{\mathcal{E}}_P$ near $T_c$ except for the small bump near 50 meV in $\mathcal{E}_N$, and (ii) on the applicability and mutual consistency of the Eliashberg equations for the normal and pairing self-energies to a similar accuracy.

**Salient Points of the Experimental Results and their Implications**

We have extracted the electron self-energies in both pairing and full lattice symmetry directly from the ARPES data without adjustable parameters using the procedure described in **SII**. We have then used the integral equations, which are shown in **SIII** to be exact, to deduce numerically both the pairing and the normal Eliashberg functions. There are no theoretical assumptions underlying our results, except that superconductivity is due to generalized BCS pairing. We summarize here the principal conclusions of our data and analysis:

(**A**) The experimental results and fits to them in Fig. 2 show that the imaginary part of the normal self-energy above, is independent of **k** and linear in $\omega$ to a good approximation as found earlier [14, 15]. The strange metal anomalies such as the linear-in-$T$ resistivity and corresponding aspects in optical conductivity follow [9]. It acquires superconductivity-induced features at low energy below $T_c$. The pairing self-energy near $T_c$ is nearly a constant as a function of $\omega$ up to the upper cut-off, and acquires the same superconductivity induced features below $T_c$, as in the normal self-energy.

(**B**) One finds on comparing in Fig. 4 that near $T_c$, $\tilde{\mathcal{E}}_P(\theta, \omega) \approx \mathcal{E}_N(\theta, \omega)$ to within the stated accuracy, except for the small bump near 50 meV in the latter. The part of these (dimensionless) quantities that is nearly a constant as a function of $\omega$ has the same magnitude as that of the normal state $\mathcal{E}_N$ and has a value of about 0.15 for UD89 and consistently somewhat smaller for OD82 up to the angle-dependent cut-off, which is about 0.2 eV at 20° moving continuously to about 0.4 eV at 45°. There are superconductivity-induced additional features in both functions as the temperature decreases well below $T_c$, and within our accuracy they have the same frequency dependence.

(**C**) Above $T_c$, $\mathcal{E}_N(\theta, \omega)$ consists of a low-energy bump at about 50 meV with a half-width of about 10 meV on top of a nearly constant part up to the angle-dependent cut-off. $\tilde{\mathcal{E}}_P(\theta, \omega)$, which can be deduced only below $T_c$, has no low energy bump near $T_c$. This means that   there is no coupling of fermions in the attractive d-wave channel to the excitations that appear in the bump but coupling to them in the s-wave channel. On the other hand the nearly constant part has a similar magnitude of coupling to fermions both in the s-wave and the d-wave channels.

(**D**) It is well understood [22, 23] that d-wave pairing, i.e. $\phi(\theta, \omega) \propto \cos 2\theta$, is favored only when fermions scatter dominantly over angles of $\pm \pi/2$. Together with the points (**C**), this exposes the central paradox of d-wave superconductivity in cuprates: The fundamental physics of the cuprates requires that the same fluctuations which dominantly scatter at angles $\pm\pi/2$ in the attractive channel must lead to a nearly angle-independent repulsive scattering in the normal channel with the full symmetry of the lattice.

It is reasonable that the bump in the normal state is due to optical phonons of the apical oxygens, as has been suggested [24]. In an interesting time-resolved conductivity experiment [21] at room temperature, the results were analyzed with fluctuations which could be divided into a peak around 50 meV and a broad electronic continuum with a cut-off at about 0.4 eV; the bump has a relaxation rate much slower than the continuum, indicating independent sources for the two contributions. Our results for $\mathcal{E}_N$ are consistent with those with the additional information through $\mathcal{E}_P$ that the feature around 50 meV has no attractive coupling in the d-wave channel.

*Determining $T_c$*

One may wish to calculate $T_c$ directly from the deduced Eliashberg functions. Such a check is however circular, since the Eliashberg functions are obtained from the solution of the equations whose linearized version gives $T_c$. Thus, if the complete information over the entire Brillouin zone were available, using the Eliashberg function near $T_c$ back in the linearized Eliashberg equations would give the experimental $T_c$. Such an exercise may however be taken as a test of several of the steps in extracting the final results from the experiments and of the extrapolations. Using the extrapolations for $\mathcal{E}_P$ from the angles measured so that their upper cut-off at other angles is also the same as the measured cut-off of $\mathcal{E}_N$ and extrapolating both from the measured angles to all angles, the linearized Eliashberg equations give for deductions using $\xi(\mathbf{k})$ of Ref. 25 that $T_c \approx 135$ K for the UD89 sample, and $T_c \approx 90$ K for the OD82 sample.

We can get an estimate of the dimensionless coupling constants in the s and d-wave channels, by using the approximate expressions for them [26, 27],

$$\lambda_s \approx\; <\int_0^\infty d\omega \frac{2}{\omega} \mathcal{E}_N(\theta,\omega)>_\theta, \qquad \lambda_d \approx\; <\int_0^\infty d\omega \frac{2}{\omega} \cos(2\theta)\mathcal{E}_P(\theta,\omega)>_\theta . \qquad (1)$$

Using the results in Fig. 4C that near $T_c$, $\mathcal{E}_N(\theta, \omega) \approx \tilde{\mathcal{E}}_P(\theta, \omega) \approx 0.15$ from about $T_c$ up to the cut-offs $\omega_c(\theta)$ and $\approx 0.15 \omega/2T$ for $\omega \lesssim 2T$ gives $\lambda_s \approx 2\lambda_d \approx 1.2$. In materials like the cuprates, where pair-breaking due to inelastic scattering is important, these parameters alone do not determine $T_c$ [27]. It is important to note, however, that the deduced fluctuations provide an enhancement $O(\ln(\omega_c/T_c))$ of the effective coupling constants which enter as crucial factors in a proper determination of $T_c$ for the kind of fluctuation spectrum deduced.

*Brief Comparison of Models for Cuprates with the Experimental Results*

The results available in the literature for calculations starting with different physical ideas are compared to the experiments in **SIV**. We summarize the comparisons here.

All calculations use adjustable parameters with which features of the experiments, such as $T_c$ may be reproduced. The comparison with experimental results must then be done with respect to the momentum and frequency dependence of the pairing and normal self-energies and Eliashberg functions noted in the principal conclusions **A − D**.

(i) The calculations [28] using measurements [29] of antiferromagnetic spin-fluctuations (in LSCO for T/Tc $\approx 0.25$) in Eliashberg theory correctly give $\phi(\theta, \omega)$ consistent with $\propto \cos(2\theta)$. The calculations give a peak in $\phi(\theta, \omega)$, reproduced in Fig. S3 of **SIV** at about 0.1 eV and nearly zero for the pairing self-energy beyond it. In Fig. (2,4) of [28], $\Sigma(\omega, \theta)$ is not linear in $\omega$ and is strongly angle-dependent. To come to conclusions about the applicability of the theory, these results should be compared with the results in Fig. 3, which has a constant part in $\phi(\theta, \omega)/\cos(2\theta)$ up to the cut-off and a linear in $\omega$ and nearly angle-independent part in $\Sigma(\omega, \theta)$ at all temperatures. Since no measurements are available at higher T, comparisons near Tc are not possible.

(ii) In the results of a very extensive dynamical mean-field theory calculations on the Hubbard model [30], [31], a value of nearest neighbor hopping parameter, t $\approx 0.3$eV, is chosen in the calculations to get nearly the right maximum value of $T_c \approx$ t/30. Only angle-averaged self-energies are calculated in this technique. The calculations give peaks in the pairing self-energy (reproduced in Fig. S4 of SIV) at energies of about 0.2t and t, i.e., between 0.06 and 0.3 eV which sharply decrease to zero in between. The imaginary part of such a $\Sigma(\omega)$ (see Fig. 5 of [31]) is constant beyond $\omega \approx 0.2$t $\approx 0.06$ eV with a peak below this value. The calculations are done for temperatures just below Tc. To come to conclusions about the applicability of the theory, these results must be compared with the experimental results in Fig. 3, which give just below Tc, a linear in $\omega$ self-energy and a nearly constant pairing self-energy up to the high frequency cut-off.

(iii) The spectra of loop-current fluctuations are calculated as the quantum-critical fluctuations of the dissipative quantum XY model. It is proportional to $\tanh(\omega/2T)$ [32] with a high frequency cut-off, which fits the deduced Eliashberg functions near $T_c$ (except for the low energy bump in $\mathcal{E}_N(\omega)$). It leadsto the well-known angle and frequency dependence of the measured normal self-energies. In Fig 5, we have shown in effect that it also gives the measured pairing self-energy, near Tc and well below Tc. We recapitulate in SIV earlier results [33] about the momentum dependence of the matrix-elements coupling the fermions to fluctuations of the model, so as to give both a repulsion in the normal channel and attraction in the pairing channel with the same frequency dependent spectra. The physics behind the central paradox thereby follows.

Any other ideas and calculations may be compared with the robust experimental results.

**Materials and Methods**

Optimally-doped and slightly under-doped Bi2Sr2CaCu2O8+_ (Bi2212) single crystals were grown by the traveling solvent floating-zone method. Over-doped Bi2212 sample was prepared by annealing the optimally-doped sample in owing oxygen atmosphere. All the samples have high quality which exhibit sharp superconducting transitions with a transition width of ~2 K.

The angle-resolved photoemission measurements were carried out on our vacuum ultra-violet (VUV) laser-based angle-resolved photoemission system [34]. The photon energy of the laser is 6.994 eV with a bandwidth of 0.26 meV and the energy resolution of the electron energy analyzer (Scienta R4000) is set at 0.5~1 meV, giving rise to an overall energy resolution of ~1.0 meV. The angular resolution is about 0.3 degree, corresponding to a momentum resolution of ~0.004 Å$^{-1}$ at the photon energy of 6.994 eV. All samples were cleaved *in situ* and measured in vacuum with a base pressure better than 5x10$^{-11}$ Torr. More details about the sample and the experiments can be found in [6,7,35].

**Acknowledgement** CMV wishes to thank Elihu Abrahams for a discussion of the exact relation of irreducible vertices to the self-energy. HYC's work was supported by National Research Foundation (NRF) of Korea through Grant No. NRF-2013R1A1A2061704. XJZ thanks financial support from the NSFC (11190022 and 11334010), the MOST of China (973 program No:

2011CB921703 and 2011CBA00110) and the Strategic Priority Research Program (B) of the Chinese Academy of Sciences (Grant No. XDB07020300). CMV's work is partially supported by NSF grant DMR 1206298. All data needed to evaluate the conclusions in the paper are present in the paper and/or the Supplementary Materials. Additional data related to this paper may be requested from the authors.

**Author Contributions**

The project was conceived by CMV, XJZ and HYC . The measurements were made by WTZ and JFH under the supervision of XJZ. The experimental data were analyzed by JFH, WTZ, YXZ, LY and JMB under the supervision of HYC, CMV and XJZ. The extraction of the self-energy and the Eliashberg Functions and the theoretical modeling were carried out by JMB and JJB, and the figures were plotted by JMB, WTZ and JFH, under the supervision of HYC, CMV and XJZ. The manuscript was written by CMV, HYC and XJZ. All authors participated in the discussion of the results and the paper writing.

**Competing Interests:** The authors declare that they have no competing interests.

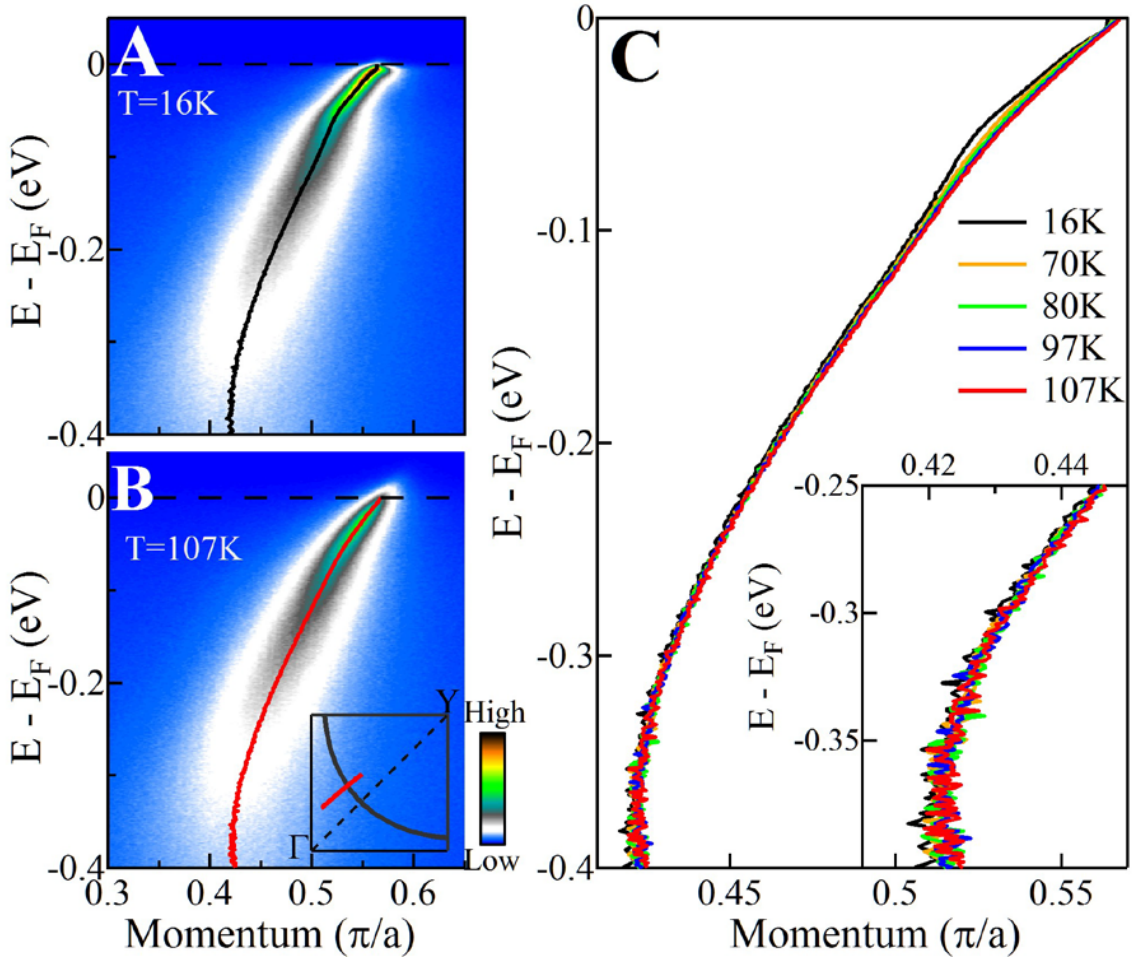

Fig. 1: **Color representation of the measured photoemission intensity** from the UD89 sample along the $\theta = 35°$ direction as shown in the inset of (b). Panel (A) is at $T = 16$ K and (B) at 107 K. Panel (C) gives the progression of the energy-momentum dispersions at temperatures 16 K, 70 K, 80 K, 97 K and 107 K. The inset in (C) gives on an expanded scale the illustration of the consistency of the data to an accuracy of $5 \times 10^{-3} (\pi/a)$ in the region at high energy where no temperature dependent corrections to the dispersion are necessary. In **S2** in Supplementary Section, we show the systematic errors in the data when such accuracy is not met and how we correct for them.

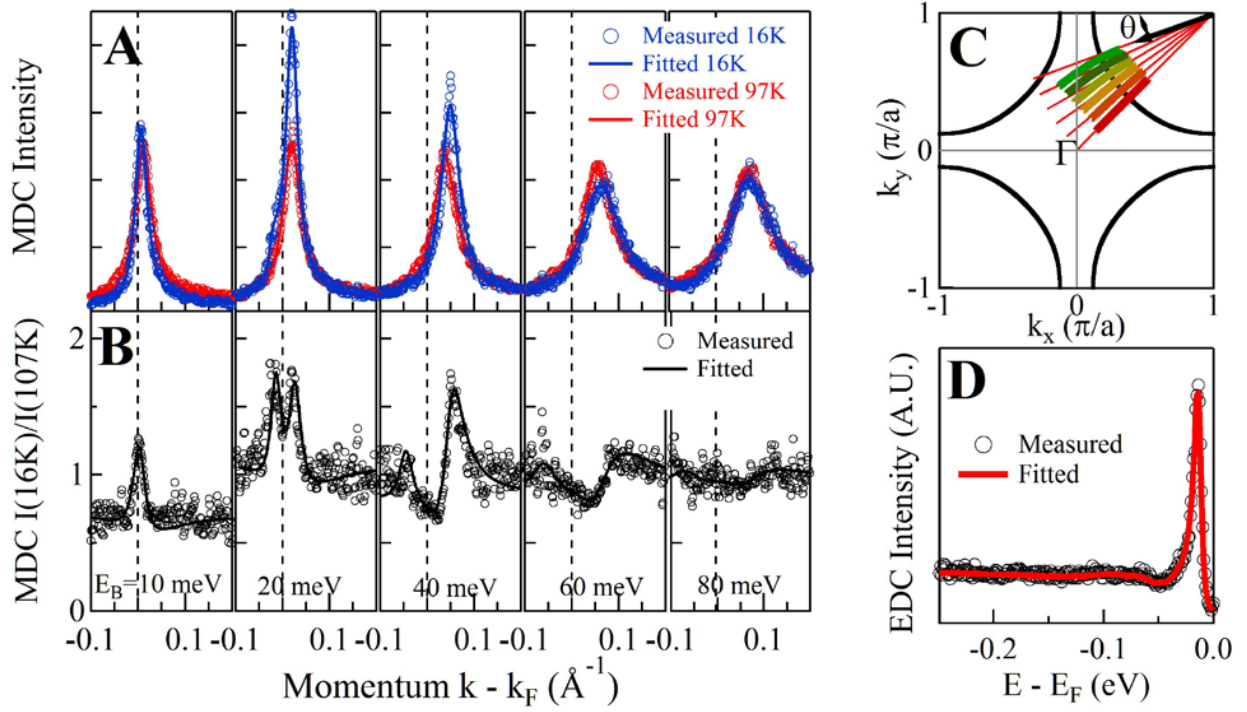

Fig. 2: **Measured MDCs and fits to them** at five different energies at 97 K and at 16 K in the UD89 sample along the dark green trajectory in C ($\theta = 20°$) are given in A. Such data were taken at 1 meV interval and at all the trajectories shown in C. The vertical scale in A is the measured photoelectron current for a fixed photon flux in arbitrary units. It is crucial in our measurements that this scale remain within the error bars discussed in the text at all temperatures, energies and momenta and any systematic errors in it be corrected. B shows the normalized difference in the measurements at the two temperatures in A and the same quantity calculated from the fits in A. The fits to the MDCs were made according to the procedure described in **S1**. The normal and the pairing self-energies are extracted from such fits, also as described in **S1**. D is intended to show how consistent are the fits to the MDC at different energies – it shows the energy distribution curve (EDC) generated from the MDC at the complete range of energies measured at an angle $\theta = 20°$ at the Fermi-surface at 16 K, and compares with the direct measurement of the EDC at the same point and same temperature.

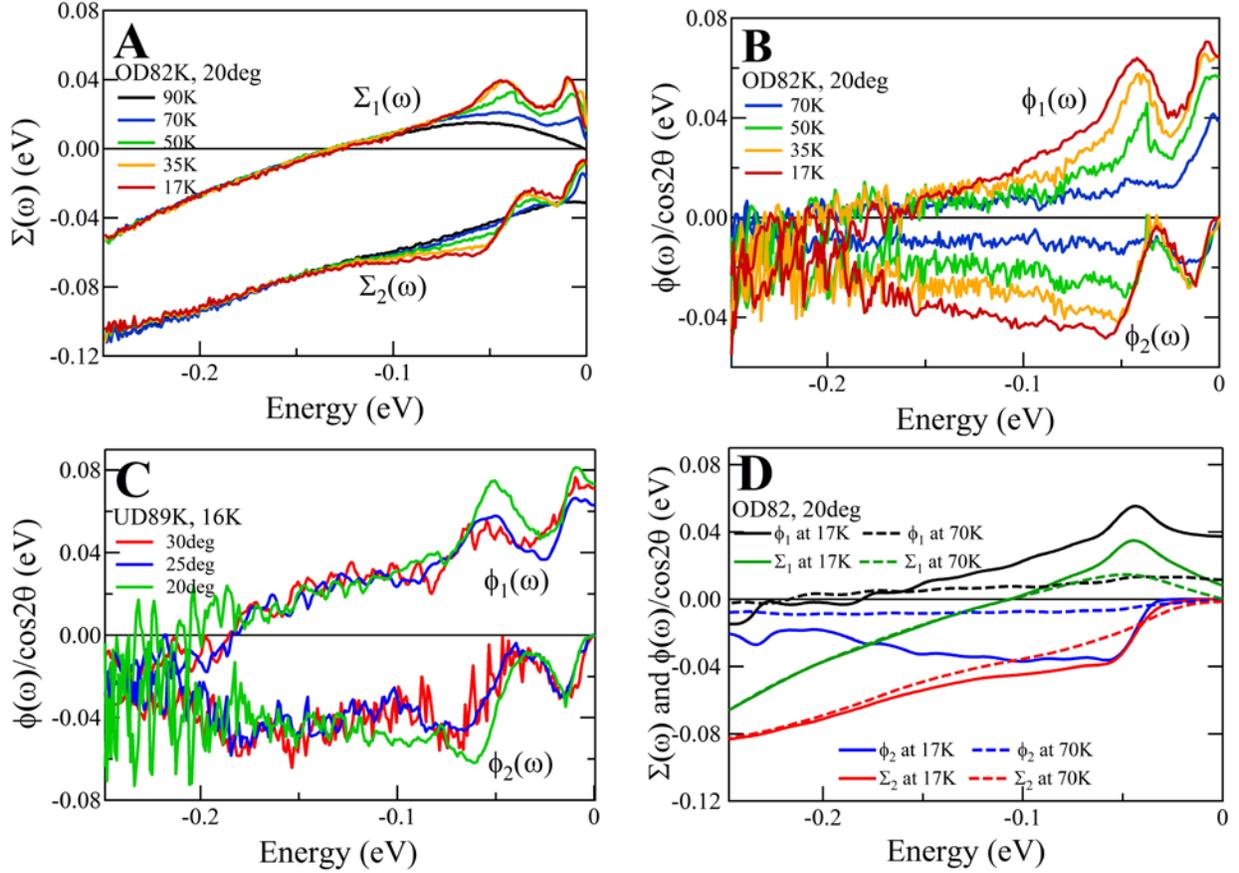

Fig. 3: **Normal and pairing self-energies.** Panel A shows the evolution of the extracted normal self-energy and B the evolution of the pairing self-energy as a function of temperature directly from the fits to the MDCs in OD82. The normal and the pairing self-energy show superconducting gap induced features at low energies up to about $3\Delta$, and are smoothly varying in energy thereafter up to a cut-off energy. Panel C shows the pairing self-energies in UD89 at 16 K divided by $\cos(2\theta)$. The determination of the pairing self-energy has acceptable signal to noise ratios till about 0.2 eV only. The data fall together at the angles shown to an accuracy of better than 10% till about 0.2 eV. Panel D shows the self-energies smoothed over $\pm5\text{meV}$ as discussed in the text and after removing the impurity induced features for OD82 at $T = 17$ K (solid lines) and 70 K (dashed lines).

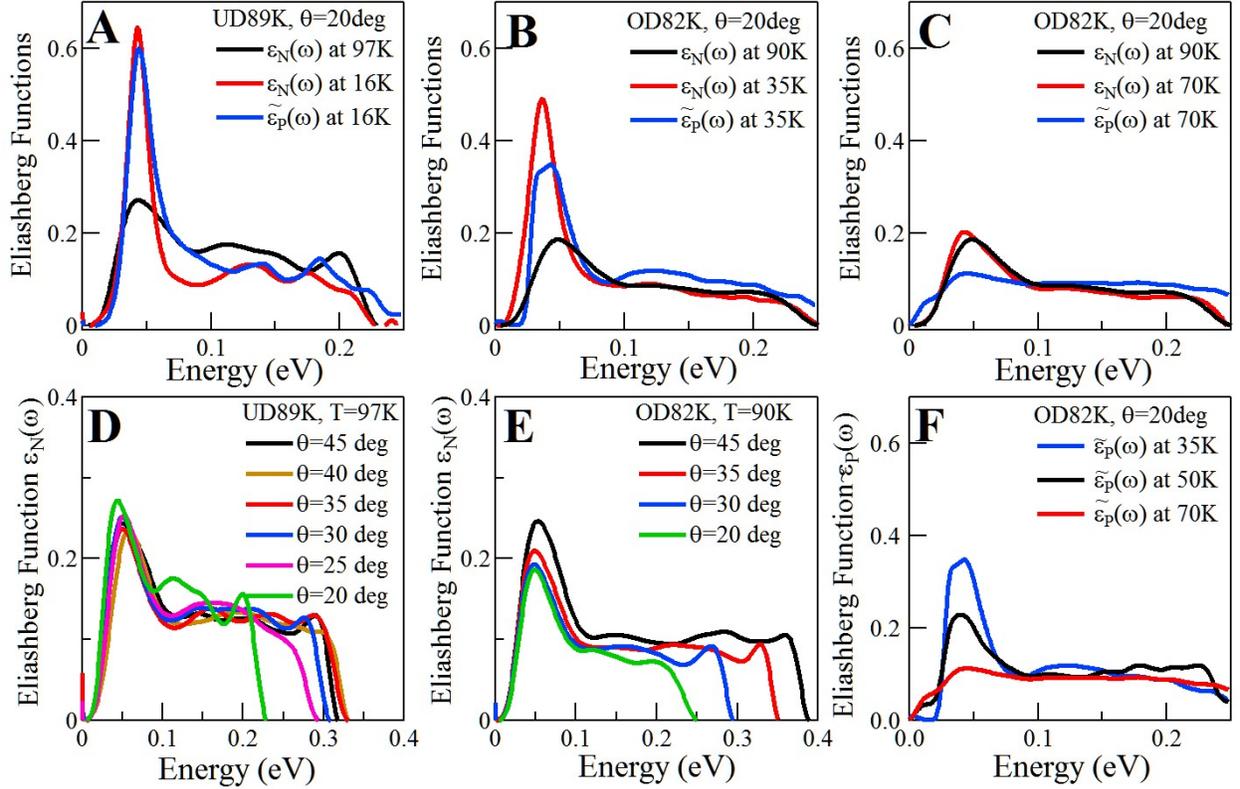

Fig. 4: **The Eliashberg functions**: Normal $\mathcal{E}_N(\theta, \omega)$, and the scaled pairing Eliashberg functions, $\tilde{\mathcal{E}}_P(\theta, \omega) \equiv \mathcal{E}_P(\theta, \omega)/\cos(2\theta)$. These are calculated by solution of the Eliashberg equations from the measured self-energies. Panel A and B compare $\tilde{\mathcal{E}}_P(20°, \omega)$ and $\mathcal{E}_N(20°, \omega)$ deep in the superconducting state for the two samples, and the latter also above $T_c$. At low temperatures, they are the same to our accuracy over the whole frequency range, with a large superconductivity induced enhanced low energy peak. Panel C shows that closer to $T_c$, the low energy peak in $\tilde{\mathcal{E}}_P(\omega)$ disappears. This trend is more directly shown in panel F. Panels D and E give $\mathcal{E}_N(\theta, \omega)$ for $T$ above $T_c$, showing the increase in the cut-off energy with increasing $\theta$. The gentle waviness in all of the results are artifacts of the maximum entropy method for the solution of the Eliashberg equations.

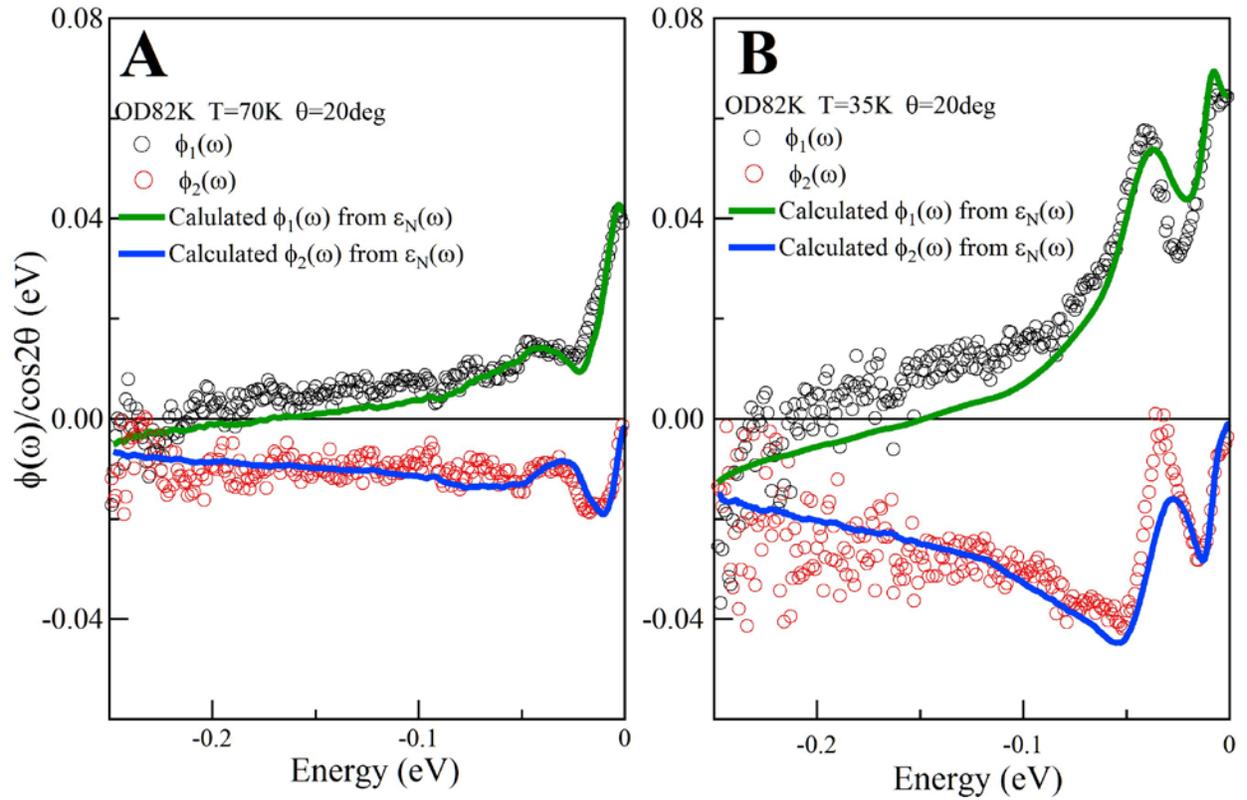

Fig. 5: **Calculation of pairing self-energies assuming** $\tilde{\mathcal{E}}_P = \mathcal{E}_N$. The experimentally deduced real and imaginary part of the self-energy in absolute units are compared with a calculation of the same quantity from the Eliashberg Equations assuming that $\tilde{\mathcal{E}}_P = \mathcal{E}_N$ for $T = 70$ K and 35 K in OD82 sample. This agreement occurs only if the Eliashberg equations are applicable for the analysis of the data and the relation of the two Eliashberg functions.

# Supplementary Information

## Quantitative Determination of the Pairing Interactions for High Temperature Superconductivity in Cuprates

**SI. Extraction of Normal and Pairing Self-Energy from ARPES in the Superconducting State**

SI.1 The Single-Particle Spectral Function

SI.2 Procedure for Extracting the Self-Energy

**SII. Correction of Systematic Errors and Renormalization of the ARPES Data**

SII.1 Limits of Validity of Results

**SIII. Equations for the Self-Energy**

SIII.1 Exact Representation of the Self-Energy

SIII.2 Familar Eliashberg Integral Equations for d-wave Superconductors

**SIV Comparison of Theories with the Results from Experiments.**

# I.   Extraction of Normal and Pairing Self-Energies from ARPES in the Superconducting State

For the physics of the ARPES process, we refer to excellent reviews [36,37]. ARPES measures, for a given flux of photons of energy $\nu$ incident on a sample, the intensity of photo-electrons of kinetic energy $E_{kin}$ and chosen angles at the detector with respect to the crystalline axes. Using the energy and momentum conservation laws, the kinetic energy and the angles can be converted into the energy $\omega$ and the crystal momentum $\mathbf{k}$ of the one-particle state of the sample before the photo-excitation. For the purposes of this work, it is necessary to measure the ARPES intensity $I(\mathbf{k}, \omega)$ at different $\mathbf{k}$ and $\omega$ of interest and temperature $T$ from just above $T_c$ to well below it, with an accuracy of better than about 2%. Fig. 1 and Fig. 2 for UD89 give an idea of the quality of the raw data. We show similar results for the OD82 sample in Fig. S1, which are of somewhat poorer quality than in Fig. 1, but quite adequate for extracting the self-energies to the accuracy necessary for our conclusions.

In this section, we first cast $I(\mathbf{k}, \omega)$ in a form suitable for our analysis in **SI.1**, and then explain the procedure of extracting the normal and pairing self-energies from the data in **SI.2**. The results obtained depend on the accuracy and consistency of the experimental data, which are checked. One encounters the issues of signal to noise in the data as well as systematic errors due to both variation of photon flux and the small movements of sample with respect to the source of photons and the detector as a function of temperature. One also needs to renormalize the momentum distribution curve (MDC) in the superconducting state such that any slight misfits in the normal state do not affect the superconducting state fits. In other words one should make sure that the pairing self-energy and accompanying deviation of the normal self-energy from above $T_c$ are extracted only from the difference between the MDC data between below and above $T_c$ for the same cut and $\omega$. We explain how we minimize and take into account the systematic errors and how to renormalize the superconducting state MDC data in **SII**.

## SI.1   The Single-Particle Spectral Function

The ARPES intensity $I(\mathbf{k}, \omega)$ for unit-incident flux of photons is given, in the sudden approximation, by

$$I(\mathbf{k}, \omega) = |M(\mathbf{k}, \nu)|^2 f(\omega)[A(\mathbf{k}, \omega) + B(\mathbf{k}, \omega)]. \tag{S1}$$

$M(\mathbf{k}, \nu)$ is the matrix element of the photo-emission process, $f(\omega)$ the Fermi distribution function, and $B(\mathbf{k}, \omega)$ is the background from the multiple scatterings of the photo-electrons, which in well done laser based ARPES is small and well characterized in MDC measurements as seen in Fig. 2. $A(\mathbf{k}, \omega)$ is the single-particle spectral function given by the imaginary part of the retarded Green's function. Our primary interest is to extract the normal self-energy $\Sigma(\mathbf{k}, \omega)$ and the pairing self-energy $\phi(\mathbf{k}, \omega)$ in terms of which the Green's function is written.

$$A(\mathbf{k}, \omega) = -\frac{1}{\pi} Im G_{11}(\mathbf{k}, \omega), \tag{S2}$$

$$\hat{G}(\mathbf{k}, \omega) = \frac{W(\mathbf{k}, \omega)\tau_0 + Y(\mathbf{k}, \omega)\tau_3 + \phi(\mathbf{k}, \omega)\tau_1}{W^2(\mathbf{k}, \omega) - Y^2(\mathbf{k}, \omega) - \phi^2(\mathbf{k}, \omega)}, \tag{S3}$$

where the subscript in Eq. (S2) represents the 11 component of the matrix Green's function $\hat{G}$, and the $\tau_i$ ($i = 0,1,2,3$) are the Pauli matrices in the Nambu space. As verified in Fig. 2, the normal self-energy and pairing self-energy depend on $k_\perp$ very weakly, and are functions of $\theta$ and $\omega$. Then

$$W(\theta, \omega) = \omega - \Sigma_0(\theta, \omega) \equiv \omega Z(\theta, \omega), \tag{S4}$$

$$Y(\mathbf{k}, \omega) = \xi(\mathbf{k}) + \Sigma_3(\theta, \omega), \tag{S5}$$

$$\phi(\theta, \omega) = Z(\theta, \omega)\Delta(\theta, \omega). \tag{S6}$$

The $\Sigma_3(\theta, \omega)$ is in principle necessary because we need to consider self-energies over a large energy region from the chemical potential, where $\xi(\mathbf{k})$ is not symmetric, and the impurity induced resonance[10,11] in the superconducting state come from potentials which are in general not particle-hole symmetric. The normal self-energy $\Sigma(\theta, \omega)$ is given by

$$\Sigma(\theta, \omega) = \Sigma_0(\theta, \omega) + \Sigma_3(\theta, \omega). \tag{S7}$$

It evolves continuously to the normal state self-energy $\Sigma(\theta, \omega)$ above $T_c$ where the distinction between $\Sigma_0(\theta, \omega)$ and $\Sigma_3(\theta, \omega)$ can not be made in the fitting process.

In this paper, we use the bare dispersion $\xi(\mathbf{k})$ given in Eqs. (3) and (11) in Ref. [25] with 6 parameters determined from very detailed fits to the band-structure calculated by density functional methods. We have also used the 4 parameters given in Ref. [17]. Although the measured Fermi-surfaces with the two band-structures are identical, the detailed dispersions differ at higher energies. The differences in the results using the two different band-structures are discussed as a source of systematic errors in **SII**.

### SI.2   Procedure for Extracting the Self-Energy

The procedure we employ here is a combination of a refinement of the MDC fitting in Ref. [38] and a real frequency implementation of the MDC area ratio approach as proposed in Ref. [8] and [39]. The MDC self-energy analysis fits the ARPES intensity $I(\mathbf{k}, \omega)$ of Eq. (S1) and (S2) for a fixed $\theta$ and $\omega$ with a chosen bare dispersion $\xi(\mathbf{k})$ as a function of $k_\perp$ to extract the $k_\perp$-independent normal self-energy $\Sigma(\theta, \omega)$ and pairing self-energy $\phi(\theta, \omega)$. The most important and unique feature of the MDC fitting is that one can determine the normal and pairing self-energies separately on an equal footing. One should notice the almost perfect MDC fittings shown in Figs. 2. This demonstrates the experimental justification of the $k_\perp$-independence of the self-energy and the matrix element $M(\mathbf{k}, \nu)$ of Eq. (S1). The $k_\perp$-independence of $M$ comes in because the spectral function $A$ of Eq. (S1) has a much sharper quasi-particle peak as a function of $k_\perp$ (at $k_\perp = k_m(\omega)$) than $M$. The $k_\perp$ dependence of $M(k_\perp)$ then can be factored out and written as a function of $\omega$.

Now, the MDC area ratio approach takes the ratio of MDC areas in the superconducting and normal states, $\mathcal{A}_S(\theta, \omega)/\mathcal{A}_N(\theta, \omega)$, and equates it with the superconducting density of states.

$$ReN(\theta, \omega) = \frac{\mathcal{A}_S(\theta, \omega)}{\mathcal{A}_N(\theta, \omega)} = Re\left[\frac{\omega}{\sqrt{\omega^2 - \Delta^2(\theta, \omega)}}\right]. \qquad (S8)$$

This relation holds for general energy dependent DOS as well provided that the bandwidth is the largest energy scale. A combination of the two methods successfully produces the normal self-energy $\Sigma(\theta, \omega)$ and pairing self-energy $\phi(\theta, \omega)$.

The fitting of experimental MDC data $I(k_\perp, \theta, \omega)$ using Eq. (S1) in the normal state returns the $\Sigma(\theta, \omega)$ straightforwardly. In the superconducting state, however, it is a subtle matter because the fitting parameter space is expanded and yet, the parameters (the self-energies) must be determined by the small difference of the MDCs between the superconducting and normal states. Therefore the fitting must be aided by other information to ensure the reliable results. This is provided by the MDC area ratio approach.

The experimental results are taken along the trajectories shown in Fig. 2C, which are not straight lines in the $(k_\perp, \theta)$ plane, especially as the anti-nodal direction is approached. But the trajectories are known very well and one can convert from the points of measurement to $(k_\perp, \theta)$. It turns out that the corrections are significant only in $k_\perp$ compared to those in $\theta$.

To calculate the area under the MDC, for the chosen $\theta$, and a wide distribution of energies, by integrating the experimental $I(k_\perp, \theta, \omega)$ over $k_\perp$, the background $B(\mathbf{k}, \omega)$ must first be subtracted. Over most of the energy range, $B(\mathbf{k}, \omega)$ is independent of $k_\perp$ to a very good approximation, as seen in Fig. 2, and therefore can be easily determined. At higher energies, above about 0.1 eV, it is weakly momentum dependent. The observed slight asymmetry of the MDC shape as a function of $k_\perp$ may be accounted for by a deviation of the bare dispersion $\xi(\mathbf{k})$ from the linear dispersion relation like the tight-binding dispersion or $k_\perp$ dependent background. We took the tight-binding dispersion for $\xi(\mathbf{k})$ and $k_\perp$-independent background, and proceed as follows.

(1) The initial calculation of the $ReN(\theta, \omega)$ begins with the MDC fittings with $\phi = 0$ for both $I_S(k_\perp, \theta, \omega)$ below $T_c$ and $I_N(k_\perp, \theta, \omega)$ above $T_c$. This returns the backgrounds $B_S(\theta, \omega)$ and $B_N(\theta, \omega)$ as well as the normal self-energies. Then the MDC areas were calculated both in the superconducting and normal states after the backgrounds were subtracted off and used to calculate the superconducting DOS from Eq. (S8). This requires that the matrix element $M$ of Eq. (S1) cancels out exactly in taking the ratio because the $k_\perp$ integral of the $A$ gives the density of states. As discussed above, the matrix element $M$ can be factored out of the $k_\perp$ integral of MDC area. Also, because $M$ is independent of the temperature, it is cancelled out in taking the ratio.

(2) Take the Kramers-Kronig transform to obtain the imaginary part of $N(\theta, \omega)$. This gives a complete information on the complex function $N(\theta, \omega)$. The complex function $\Delta(\theta, \omega)$ is obtained from the relation,

$$\Delta(\theta, \omega) = \omega \left[ 1 - \frac{1}{N^2(\theta, \omega)} \right]^{\frac{1}{2}}. \tag{S9}$$

The self-energies $\phi(\theta, \omega)$ are obtained from

$$\phi(\theta, \omega) = \Delta(\theta, \omega) Z(\theta, \omega), \tag{S10}$$

$$Z(\theta, \omega) = 1 - \frac{\Sigma(\theta, \omega)}{\omega}, \tag{S11}$$

where $\Sigma(\theta, \omega)$ was already obtained as described in the step (1) above.

(3) Now, go back to the step (1) to make the MDC fittings of $I_S$ and $I_N$, but allowing $\phi \neq 0$ below $T_c$. The predetermined $\Sigma$ and $\phi$ in the step (2) serve as a guide to the subtle MDC fitting in the superconducting state. This returns improved $\Sigma$ and $\phi$ as well as the background $B$.

(4) Go back to the step (2) to recalculate the MDC area with the newly determined background $B$. The resulting $\Sigma$ and $\phi$ from Eq. (S10) and (S11) serve to make next iteration of MDC fittings. We iterate this process until the pairing self-energy from the MDC fitting and the MCD area ratio converge.

The self-energies presented here were obtained by the iterative process of the MDC fitting and MDC area ratio approach just explained.

  A comment should be made here on the classic work of McMillan and Rowell[1]. They measured the ratio of the density of states in a tunneling conductance experiment on Pb as a function of energy in the superconducting state to that in the state just above $T_c$. This was used to obtain the gap function $\Delta(\omega)$ from which they deduced the Eliashberg function $\mathcal{E}_P$, (called $\alpha^2 F(\omega)$ by them). This procedure works well for s-wave superconductors, where $\mathcal{E}_P = \mathcal{E}_N$, but this is not suitable for d-wave superconductors because $\mathcal{E}_P$ is in general different from $\mathcal{E}_N$. d-wave superconductivity requires two experimentally determined functions which may be determined from the coupled equations for the normal and the pairing self-energies to determine two distinct Eliahsberg functions $\mathcal{E}_N(\theta, \omega)$ and $\mathcal{E}_P(\theta, \omega)$ as is explained in **SIII**.

## II.    Correction of Systematic Errors and Renormalization of the ARPES Data

  During the measurements, the sample orientation sometimes shows a small change with temperature, originating from the thermal expansion of the connection parts of the cryostat. This drift may induce a small angle (momentum) shift in the measured photoemission spectrum. This effect is too small to be removed in situ during the measurement by realigning the sample at each temperature because of the limited motor-driven angular precision. For the high-precision data analysis performed in the present work which requires absolute measurements of counts of flux of electrons at the detector, such effects must be taken care of. We needed to carry out small intensity renormalization and angle shift corrections on the measured data. The corrections are based on the fact that, for ARPES data taken along a given momentum cut at different temperatures, the high binding energy part should show negligible change with temperature. In some of the measurements, no corrections are needed. This is illustrated in Fig. 1 in the main paper and especially the expanded panel C in it. In this case, the measurement condition is stable (negligible laser photon flux variation and negligible sample shift with temperature change), the high energy part stays the same for different temperatures in terms of both intensity and extracted dispersion (between $-0.4$ eV and $-0.3$ eV). Here no correction is necessary and MDCs are directly extracted. Fig. S1 illustrates a case in which corrections are needed. Momentum correction is performed to make sure that, at each temperature, the high energy dispersions coincide with each other by putting a small offset to the momentum along the cut direction. The coincidence of the high energy dispersions after such a correction in Fig. S1C, limited only by the noise in the data, validates the application of our momentum correction procedure.

The effect of the slight laser photon flux fluctuation on the data can be removed by normalizing the measured data at different temperatures so that the intensity of the high binding part is the same. Such an intensity normalization is performed as the first step in our data analysis.

Also important is to renormalize the MDCs in the superconducting state such that the pairing self-energy and accompanying deviation of the normal self-energy from above $T_c$ is only determined by the difference in the ARPES intensities between the superconducting and normal states. Any misfits in the normal state may affect the superconducting state fittings and cause spurious results if done without the renormalization. The sources of the misfits in the normal state are most likely from the uncertainty of the bare dispersion as discussed below. The renormalizations were done as follows: We first fit the MDCs slightly above $T_c$, i.e., 97 K for UD89 data and 90 K for OD82 data, as a function of $k_\perp$ with the enforcement of $\phi = 0$ and determine the normal state fitting curve. Then, we divide the MDC data to calculate the ratio, $I_S(k_\perp, \theta, \omega)/I_N(k_\perp, \theta, \omega)$, and multiply the ratio by the normal state fitting curve. This is the renormalized MDCs in the superconducting states we fit.

A significant source of the systematic error is the lack of precise knowledge of the bare dispersion $\xi(\mathbf{k})$ at high energies. The Fermi-surface is well fitted by more than one form of dispersion. As mentioned we have used two different parameterization of the band-structure. The results for normal self-energy near the Fermi-energy are always the same but differ far from it for $\omega$ larger than about $3\Delta \approx 65$ meV. However, we find very little variation in the pairing self-energy. This is because it is obtained from the differences of the data in the normal and superconducting states as discussed above, both of which are deduced with the same $\xi(\mathbf{k})$. The difference in the normal self-energy with the two different band-structures leads to differences which are discussed below.

The maximum entropy method for solution of integral equations can introduce unphysical oscillations in the results. These oscillations are uncontrollable in the solution of the integral equations if we take as input the raw deduced values of the self-energies, such as shown in Fig. 3. We average the measured self-energy at each energy over $\pm 5$ meV around it as inputs to the integral equations. Even with such averaging, we obtain smoothly varying oscillatory results, varying at any energy by about $\pm 10\%$, through different constraints imposed in the process of the solution. We have guided ourselves by consistency and smooth variations of the results from one temperature to another. The final results presented for the Eliashberg functions are similar to the errors in the experiments discussed above. Adding all errors in quadrature, the results may be trusted only to about 10% at any energy up to about 0.1 eV and only up to about 15% at the maximum energies of about 0.2 eV.

### SII.1 Limits of Validity of Results

Using a band-structure $\xi(\mathbf{k})$ given in high quality band-structure calculations, we can extract the absolute value of the normal self-energy $\Sigma(\mathbf{k}, \omega)$ to an accuracy of about 2% over the whole range of measurements and at all angles up to 0.45 eV. The poorest signal to noise ratio occurs in determining the pairing self-energy $\phi(\mathbf{k}, \omega)$ because it can only be extracted from the difference between the normal and superconducting state signals: $I_S(\mathbf{k}, \omega) - I_N(\mathbf{k}, \omega)$. It will be apparent below that the accuracy in extracting $\phi(\mathbf{k}, \omega)$ is better than ~10% till up to ~0.2 eV for $\theta$ between 20° and 35° but progressively gets worse at larger energies; the data are not useful to directly deduce the pairing self-energy above about 0.2 eV. Similarly, signal to noise in $\phi$ becomes poorer when the momentum cuts come closer to the diagonal direction ($\theta = 45°$) and the temperature comes closer to $T_c$, because $\phi$ gets smaller. Therefore, such data were not used in the present analysis. It is expected that the next generation ARPES apparatus will be able to alleviate these limitations. However, it is expected that the principal conclusions of this paper, using the measured results and reasoned extrapolations from it, will continue to hold.

## III.   Equations for the Self-Energies

Eliashberg derived the integral equations for the normal and the pairing self-energies starting from a Hamiltonian of electrons and phonons through the leading order perturbation in the electron-phonon interactions. This is justified by the Migdal theorem and the small magnitude typically of the parameter $\lambda \omega_D / E_F$; $\lambda$ is the dimensionless coupling constant, $\omega_D$ the Debye frequency, and $E_F$ the typical band-width. In our case, the cut-off frequency of the fluctuations is similar to the band-width and the coupling constant is about 0.5. Therefore the accuracy of the extracted Eliashberg functions from the measured self-energies may be open to question. We first show here that, given an experimentally obtained self-energy function, the momentum and energy dependence of the collective modes leading to the self-energy can be determined from the Eliashberg equations without the Migdal or weak-coupling assumptions. This is true of the McMillan-Rowell type of results also. But in that case, Migdal's theorem obviates the need to pose the question. The procedure using the experimental self-energies to deduce the fluctuations, is quite different from calculating the self-energy using a spectra of fluctuations not obeying the Migdal approximation, which may be impossibly hard.

In the normal state, the self-energies can be expressed [40,41] exactly in terms of the irreducible vertex $\mathcal{I}(\mathbf{k}, \mathbf{k}', \omega, \omega', \mathbf{q} = 0, \Omega = 0)$ and the exact Green's function $\hat{G}(\mathbf{k}, \omega)$ of Eq. (3). This is easily generalized to the superconducting state in which the relation between the self-energies, the vertices and the Green's function is shown in Fig. S2A. Assuming that the collective mode contributions to the vertex are a function primarily of the energy transfer $(\omega - \omega')$, the integral Eq. S2A for the self-energy is,

$$\hat{\Sigma}(\mathbf{k}, \omega) = \int d\omega' Tr \sum_{k'} \mathcal{I}(\mathbf{k}, \mathbf{k}', \omega - \omega', \mathbf{q} = 0, \Omega = 0) \hat{G}(\mathbf{k}', \omega'). \qquad (S12)$$

This equation is in the Gorkov-Nambu space,

$$\hat{\Sigma}(\mathbf{k}, \omega) \equiv \Sigma_0(\mathbf{k}, \omega)\tau_0 + \Sigma_3(\mathbf{k}, \omega)\tau_3 + \phi(\mathbf{k}, \omega)\tau_1. \qquad (S13)$$

Similarly $\hat{G}(\mathbf{k}, \omega)$ is given in $\tau$-space as in Eq. (S3). $\mathcal{I}$ is similarly the sum of the normal $\tau_3\tau_3$ part, $\mathcal{I}_{33}$, and a $\tau_1\tau_1$ part $\mathcal{I}_{11}$. The trace in Eq. (S12) is in $\tau$-space. $\mathcal{I}$ is irreducible in the $(\mathbf{q}, \Omega)$ particle-hole channel. Further, $\mathcal{I} = \mathcal{I}^{(0)} + \mathcal{I}^{(1)}$, the sum of a part with total spin 0 and with 1. Given the correct $\hat{\Sigma}(\mathbf{k}, \omega)$, the solution of this integral equation gives the equally correct irreducible vertex $\mathcal{I}(\mathbf{k}, \mathbf{k}', \omega, \omega')$. Since Eq. (S12) is exact, it includes all vertex corrections and self-energy insertions in a perturbative calculation of the self-energy.

One may further write, for a square lattice with s or d-wave pairing,

$$\mathcal{I}_{33}(\mathbf{k}, \mathbf{k}', \nu) = g_0 A_{1g}(\hat{k}) A_{1g}(\hat{k}') I_{33}(k, k', \nu) + .. \tag{S14}$$

$$\mathcal{I}_{11}(\mathbf{k}, \mathbf{k}', \nu) = g_{2,1} B_{1g}(\hat{k}) B_{1g}(\hat{k}') I_{11}(k, k', \nu)$$
$$+ g_{2,2} B_{2g}(\hat{k}) B_{2g}(\hat{k}') I_{11}(k, k', \nu) + ... \tag{S15}$$

where $A_{1g}, B_{1g}, B_{2g}, ..$ are the relevant irreducible representations of the point group and the $g$'s are the corresponding coupling constants.

Since the normal self-energy must be of $A_{1g}$ symmetry and superconductivity in cuprates is in $B_{1g}$ symmetry, it follows from Eq. (S12) that measurement of the normal and pairing self-energy and the solution of that equation yields, on integration over $\mathbf{k}'$, (using symmetry of the vertices under interchange of $k$ and $k'$), the irreducible vertices in the normal $I_{33}(k, k', \nu)$ and the pairing $I_{11}(k, k', \nu)$ channels. Eqs. (S12, vertex-decomp) are shown below to be identical to the more familar Eliashberg equations.

### SIII.2 Familar Eliashberg Integral Equations for d-wave Superconductors

If as assumed above, the dependence of the irreducible vertex on $\omega$ and $\omega'$ is only through the energy transfer $(\omega - \omega')$, Fig. S2A are identical to the more familiar skeleton diagrams for the self-energy, shown as Fig. S2B. To show this, we identify that the irreducible vertex for the normal self-energy $\mathcal{I}_{33}(\mathbf{k}, \mathbf{k}', \nu) = g(\hat{k}, \hat{k}') F(\mathbf{k}, \mathbf{k}', \nu) g(\hat{k}', \hat{k})$, and for the pairing self-energy $\mathcal{I}_{11}(\mathbf{k}, \mathbf{k}', \nu) = g(\hat{k}, \hat{k}') F(\mathbf{k}, \mathbf{k}', \nu) g(-\hat{k}, -\hat{k}')$. Now consider Eq. (S12). On re-expressing $G(\mathbf{k}, \omega)$ in terms of the spectral function and re-arranging, the familiar Eliashberg equations [2,3] (S16-S17) below, as generalized for d-wave superconductivity[27,13], follow in terms of the normal Eliashberg function $\mathcal{E}_N(\theta, \omega)$ and the d-wave pairing Eliashberg function $\mathcal{E}_P(\theta, \omega)$, defined below.

$$\Sigma(\theta, \omega) = \int_{-\infty}^{\infty} d\omega' L(\omega, \omega') \mathcal{E}_N(\theta, \omega'), \tag{S16}$$

$$\phi(\theta, \omega) = -\int_{-\infty}^{\infty} d\omega' M(\omega, \omega') \mathcal{E}_P(\theta, \omega'), \tag{S17}$$

$$L(\omega, \omega') \equiv \int_{-\infty}^{\infty} \frac{f(\varepsilon) + n(-\omega')}{\varepsilon + \omega' - \omega - i\delta} N_1(\varepsilon). \tag{S18}$$

$$M(\omega, \omega') \equiv \int_{-\infty}^{\infty} \frac{f(\varepsilon) + n(-\omega')}{\varepsilon + \omega' - \omega - i\delta} D_1(\varepsilon). \tag{S19}$$

$$N_1(\varepsilon) \equiv < Re \frac{W(\theta', \varepsilon)}{\sqrt{W^2(\theta', \varepsilon) - \phi^2(\theta', \varepsilon)}} >_{\theta'} \tag{S20}$$

$$D_1(\varepsilon) \equiv < Re \frac{2 \phi(\theta', \varepsilon) \cos(2\theta')}{\sqrt{W^2(\theta', \varepsilon) - \phi^2(\theta', \varepsilon)}} >_{\theta'} \tag{S21}$$

Here $<..>_{\theta'}$ implies the normalized integral over $\theta'$. The normal and pairing Eliashberg functions are given in terms of the coupling constant and interaction vertices,

$$N_1(\varepsilon)\mathcal{E}_N(\mathbf{k}, \omega) = \int d\mathbf{k}' A_N(\mathbf{k}', \varepsilon)[|g^{(0)}(\mathbf{k}, \mathbf{k}')|^2(-\frac{1}{\pi})Im\mathcal{F}^{(0)}(\mathbf{k}, \mathbf{k}', \omega)$$
$$+ 3 \left| g^{(1)}(\mathbf{k}, \mathbf{k}')|^2 \left(-\frac{1}{\pi}\right) Im F^{(1)}(\mathbf{k}, \mathbf{k}', \omega) \right], \tag{S22}$$

$$D_1(\varepsilon)\mathcal{E}_P(\mathbf{k}, \omega) = \int d\mathbf{k}' A_\phi(\mathbf{k}', \varepsilon)[g^{(0)}(\mathbf{k}, \mathbf{k}')g^{(0)}(-\mathbf{k}, -\mathbf{k}')(-\frac{1}{\pi})Im\mathcal{F}^{(0)}(\mathbf{k}, \mathbf{k}', \omega)$$
$$- 3g^{(1)}(\mathbf{k}, \mathbf{k}')g^{(1)}(-\mathbf{k}, -\mathbf{k}')(-\frac{1}{\pi})Im F^{(1)}(\mathbf{k}, \mathbf{k}', \omega)] \tag{S23}$$

$F^{(0)}(\mathbf{k}, \mathbf{k}', \omega)$ and $F^{(1)}(\mathbf{k}, \mathbf{k}', \omega)$ are, respectively, the spin-0 and spin-1 fluctuation propagators. Their corresponding vertices with fermions are $g^{(0)}(\mathbf{k}, \mathbf{k}')$ and $g^{(1)}(\mathbf{k}, \mathbf{k}')$ respectively. $A$ and $A_\phi$ are, respectively, the normal single-particle spectral function of Eq. (S2) and the pairing part (12 component in $\tau$-space) of the matrix single-particle spectral function given in terms of Eq. (S3). $A(\mathbf{k}, \omega)$ has the full symmetry of the lattice $A_{1g}(\hat{k})$. And $A_\phi(\mathbf{k}, \omega)$ has the angular dependence $B_{1g}(\hat{k})$. In the isotropic approximation, the angle-dependences are 1 and $\sqrt{2}\cos(2\theta)$, respectively.

Eqs. (S16 – S23) are used to solve for the Eliashberg functions. We first calculate $N_1$ and $D_1$ from Eq. (S20) using the self-energies extracted from the fits to MDCs, and then calculate the matrices $M$ and $L$ from Eq. (S18 – S19). The inverse matrices of $L$ and $M$ are used to obtain $\mathcal{E}_N$ from $\Sigma$ and $\mathcal{E}_P$ from $\phi$. The inversion matrices $L^{-1}$ and $M^{-1}$ were calculated using the maximum entropy method to treat the logarithmically singular cases as well.

# IV.  Comparison of theories with the results from experiments

We now consider some of the ideas and calculations for models of cuprates in relation to the Eliashberg theory and the experimental results presented in the main paper.

## *Phonons*

The only part of the effective interactions deduced by the experiments in the characteristic energy range of the phonons is the broad feature around 50 meV. But this is absent in the pairing Eliashberg function $\mathcal{E}_P$ in the d-wave channel near $T_c$. As generally agreed, phonons are not responsible for d-wave pairing in the cuprates.

## *Antiferromagnetic Fluctuations*

In the usual theory[22,42] of promotion of d-wave superconductivity by antiferromagnetic (AFM) fluctuations, $g^{(1)}(\mathbf{k}, \mathbf{k}')g^{(1)}(-\mathbf{k}, -\mathbf{k}') = |g^{(1)}(\mathbf{k}, \mathbf{k}')|^2$. The spin-1 fluctuations $\mathcal{F}^{(1)}(\mathbf{k}, \mathbf{k}', \omega)$, projected to the $B_{1g}$ channel leads to a sign for $\mathcal{E}_P$ which is opposite to that projected to identity. d-wave superconductivity is therefore expected. The AFM fluctuations of the Hubbard model do provide an attractive pairing in the d-wave channel if their correlation length is much larger[22,42,43] than $k_F^{-1}$.

Systematic calculations[28] have used the measured $q$-dependent spin-fluctuation spectra[29] in La$_{2-x}$Sr$_x$CuO$_4$ at various $x$ in the Eliashberg equations to calculate the momentum and frequency dependence of the normal and pairing self-energies, $\Sigma(\mathbf{k}, \omega)$ and $\phi(\mathbf{k}, \omega)$. The unknown coupling constant is adjusted to give the measured values of $T_c$. This coupling constant has to be adjusted upwards with increased doping because the amplitude of the fluctuations go down with doping in the measurements while $T_c$ goes up. The calculated results for the pairing self-energy near optimal doping at various angles are shown in Fig. S3. The angle dependence shows the $B_{1g}$ dependence of the d-wave gap, but the frequency dependence is strongly peaked (with some structure in between) in the low energy region at about 0.1 eV and then goes to 0 rapidly. The physics of the peak and the reason for the rapid vanishing of the self-energy are fully discussed in Ref. [28] in terms of the measured antiferromagnetic fluctuations. The dependence in Fig. S3 should be compared with the experimental results for the pairing self-energy shown in Fig. 3, where a nearly constant dependence is found up to the cut-off, beside the superconductivity induced low energy features. Similar results with peaking of the pairing self-energy at lower energy, consistent with a longer AFM correlation length, appear in the calculated results Ref. [28] on using the data at lower dopings. The calculated normal self-energy $\Sigma(\theta, \omega)$ (See Figs. (2-4) of Ref. [28]) shows a strong $\theta$ dependence and does not have a linear in $\omega$ dependence. Since the normal self-energy in these calculations is angle dependent, ideas based on such calculations cannot be used to address the "central paradox" – that the normal self-energy is angle independent while the pairing is in the d-wave channel.

*Fluctuations of the Hubbard Model*

A very sophisticated calculation of the normal self-energy and the gap $\Delta(\omega)$ starting with the one band Hubbard model, using 8 site cluster dynamical mean-field theory at various dopings has been performed[30]. Such calculations do not provide the angle dependence of the self-energy and one must assume, quite reasonably, that it is in the d-wave channel just like the experiments. In other calculations on the same model with variants of the same technique[44], d-wave superconductivity is found. We show in Fig. S4 the frequency dependence of the gap $Im\Delta(\omega)$. These are also peaked at low energies, at about 0.2 and 1 in units of the kinetic energy parameter $t$, with nearly 0 in between. The parameter $t$ is adjusted in these calculations to be about 0.3 eV because the maximum $T_c$ as a function of doping in the calculations is $\approx t/60$. The normal self-energy has also been calculated, for example Fig. (5) in Ref. [31]. The normal self-energy is constant beyond about 0.2 $t$. The calculated results should be compared with the experimental results in Fig. 3.

*Fluctuations of Loop-Current Order*

The motivation of this model comes from the observation of diffraction pattern with polarized elastic neutron scattering[45] consistent with loop-current order[46,47] in four different families of underdoped cuprates, with a quantum-critical point near optimal doping. The quantum-critical fluctuations of the loop-current model belong to the universality class of the dissipative quantum XY model. Over a range of parameters, they have been derived[48,49] and checked and extended by Monte-Carlo calculations[32.] Such fluctuations provide $\mathcal{F}_0(\mathbf{k}, \mathbf{k}', \omega)$ with a scale-invariant frequency dependence $\propto \tanh(\omega/2T)$ in the normal state up to a cut-off $\omega_c$. This frequency dependence is consistent with the frequency dependence of $\mathcal{F}_0(\omega)$ deduced above from $\mathcal{E}_N$ for $T \gtrsim T_c$, except for the 50 meV bump, and $\mathcal{E}_P$ for $T \to T_c$, as well as for the normal state singular-Fermi-liquid properties[50]. This form of $\mathcal{F}_0(\mathbf{k}, \mathbf{k}', \omega)$ leads to frequency dependence observed in the experimental normal and pairing self-energies. Evidence of the latter is the calculation and comparison with the pairing self-energy starting from the measured $\mathcal{E}_N(\omega) \approx \mathcal{E}_P(\theta, \omega)/\cos(2\theta)$ near $T_c$, given in Fig. (5).

From the point of the experimental results in this paper, a crucial aspect of such fluctuations is that they resolve the central paradox of the high $T_c$ problem in cuprates, that the normal self-energy is nearly angle-independent and the pairing is in the d-wave channel. To see this, one must consider the vertex $g(\mathbf{k}, \mathbf{k}')$ for such fluctuations to fermions. It has been derived[S32] that scattering of fermions occurs with such fluctuations through their angular momentum. For an isotropic approximation to the fermion dispersion near the Fermi-surface (results with lattice symmetry have been given in Ref. [33] but within the accuracy of the results obtained here, the isotropic approximation is adequate),

$$g(\mathbf{k}, \mathbf{k}') = i g_0 (\hat{\mathbf{k}} \times \hat{\mathbf{k}}').\tag{S24}$$

The angle-dependence in $\Sigma(\theta, \omega)$ and $\phi(\theta, \omega)$ comes from integrating the vertex part of the effective interactions,

$$|g(\mathbf{k}, \mathbf{k}')|^2 = -g(\mathbf{k}, \mathbf{k}')g(-\mathbf{k}, -\mathbf{k}') = \frac{g_0^2}{2}[1 - (\cos2\theta\cos2\theta' + \sin2\theta\sin2\theta')],\tag{S25}$$

over $\mathbf{k}'$ projected over the intermediate Green's function shown in Fig. (S2) A and B respectively. The intermediate state has the full symmetry of the lattice; only the first term in Eq. (S25) then contributes on integration over $\theta'$. One therefore finds that $\Sigma(\mathbf{k}, \omega)$ is only a function of $\omega$.

Now, consider $\phi(\theta, \omega)$. Since the intermediate state at $\theta'$ is $\propto \cos(2\theta')$, only the second term in Eq. (S25) contributes on integration over $\theta'$, so that $\phi(\theta, \omega) \propto \cos(2\theta)$. Note that this part of the vertex is attractive while the s-wave part is repulsive in the pairing channel. The central paradox of high temperature superconductivity in cuprates is thus resolved. This has required $\mathcal{F}$ to be nearly momentum-independent or equivalently be a separable function of momentum transfer and energy, as in the criticality derived for the observed order. The result that $\tilde{\mathcal{E}}_P(\omega) \approx \mathcal{E}_N(\omega)$ also follows. Corrections due to lattice symmetry may actually be expected in their ratio but not in the frequency dependence. In the isotropic approximation, the two *attractive* d-wave channels in 2D in Eq. (S25) give degenerate results. But, in the cuprates, the density of states projected to $\cos(2\theta)$ or $d_{x^2-y^2}$ symmetry is larger than that in $\sin(2\theta)$ or $d_{xy}$-symmetry favoring the former. The central paradox as to how the normal self-energy is angle-independent but the pairing self-energy has $\cos(2\theta)$ dependence is therefor resolved if the primary interaction among the fermions is through exchange of quantum critical fluctuations of the loop current order with a vertex with the symmetry of Eq. (S24).

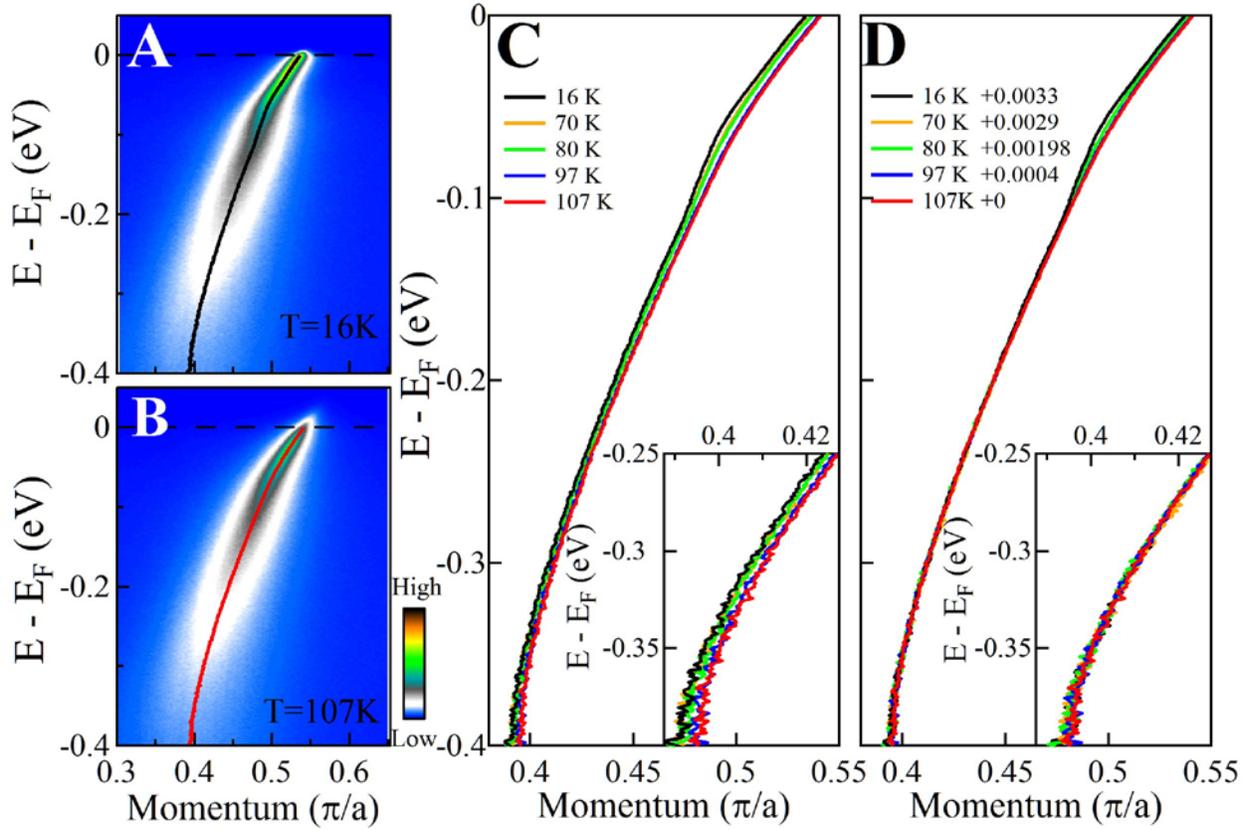

FIG. S1: **Illustration of corrections to data needed due to systematic errors due to movement of sample with change in temperature**. Color representation of the measured photoemission intensity along 10° to the Brillouin Zone in UD89 sample. (a) at 107 K and (b) at 16 K. (c) gives the progression of the Energy-momentum dispersions at temperatures 16 K, 70 K, 80 K, 97 K and 107 K. The inset in (c) gives on an expanded scale the illustration of the errors in the data due to sample movement on an expanded scale. In (d) and its inset, we show how we correct the systematic errors by aligning the high energy parts at different temperatures. The error in the raw data shown is the maximum in the data that we chose to analyze.

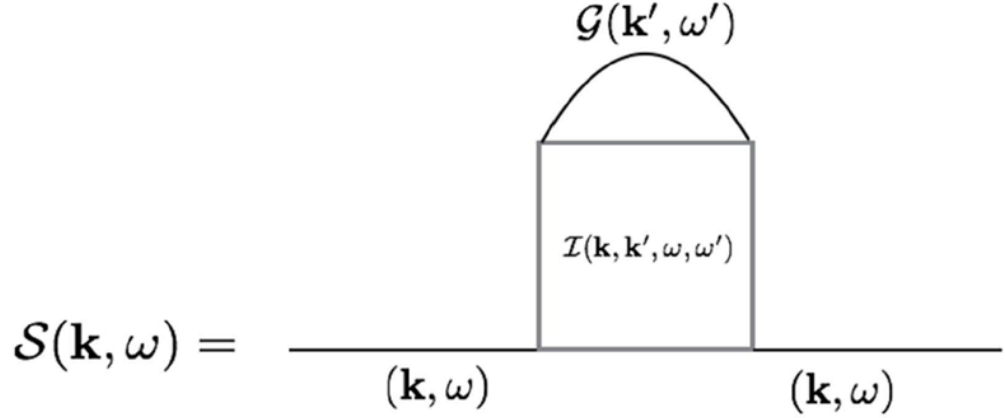

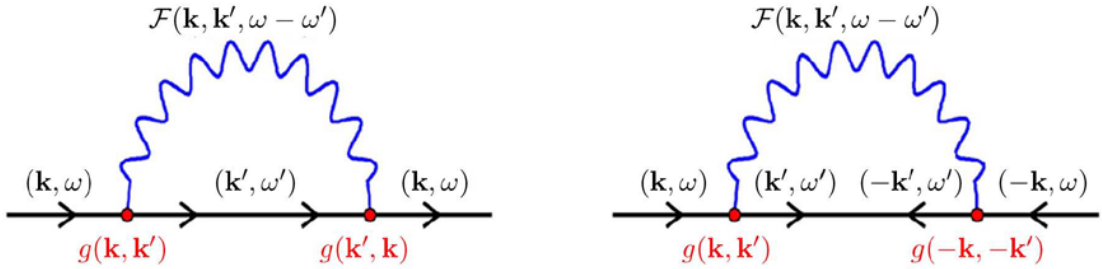

FIG. S2: **Exact representation of the normal and pairing self-energies**. **A** gives the self-energy in Gor'kov-Nambu space in terms of the exact Irreducible vertex and the exact Green's function. **B** is equivalent to **A** in terms of the more familiar skeleton diagrams, and gives on the left the the normal self-energy $\Sigma(\mathbf{k}, \omega)$ and on the right the pairing self-energy $\phi(\mathbf{k}, \omega)$. The direction of the external legs of the vertex $\mathcal{I}$, of the self-energy $\mathcal{S}$, and of the Green's function $\mathcal{G}$ in **A** for the normal and the pairing self-energy components are identical to those in **B**. The wiggly line in **B** are the fluctuations $F(\mathbf{k}, \mathbf{k}', \omega - \omega')$ exchanged by the fermions through vertices $g(\mathbf{k}, \mathbf{k}')$. The normal self-energy as a function of $(\mathbf{k})$ as well as the intermediate normal state propagator on the left as a function of $(\mathbf{k}')$ have the full symmetry of the lattice, while on the right, the intermediate pairing propagator at $\mathbf{k}', -\mathbf{k}'$ as well as the pairing self-energy at $\mathbf{k}, -\mathbf{k}$ have the symmetry of d-wave superconductivity; for example, the latter transforms as $\cos 2\theta_{\hat{\mathbf{k}}}$ in the continuum approximation.

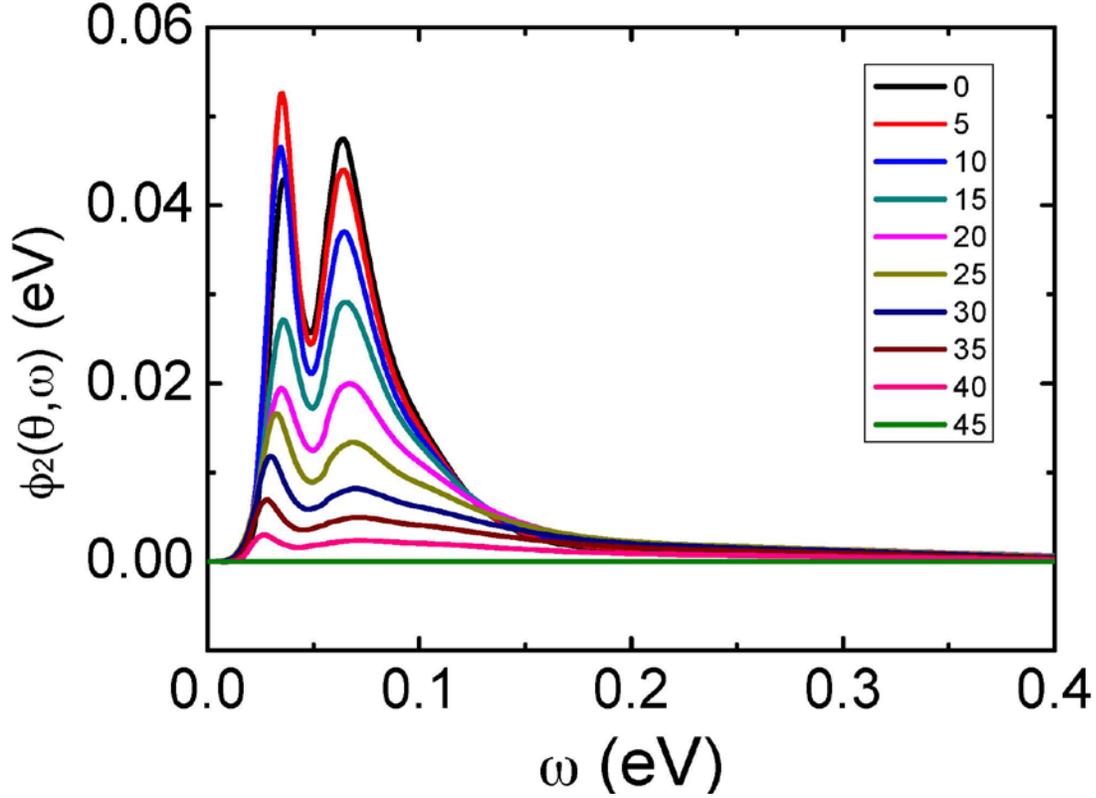

FIG. S3: **The Imaginary part of the pairing self-energy** at various angles across the Fermi-surface calculated in Ref. [28] from the measured spin-fluctuation spectra in $La_{2-x}Sr_xCuO_4$ by Vignolle et al. Ref. [29] at optimal doping. Results at other dopings may be read in the references given as also the calculated normal self-energy. The frequency dependence calculated does not compare well with the experimental pairing self-energy in Fig. 3, although its angular dependence is consistent with d-wave superconductivity.

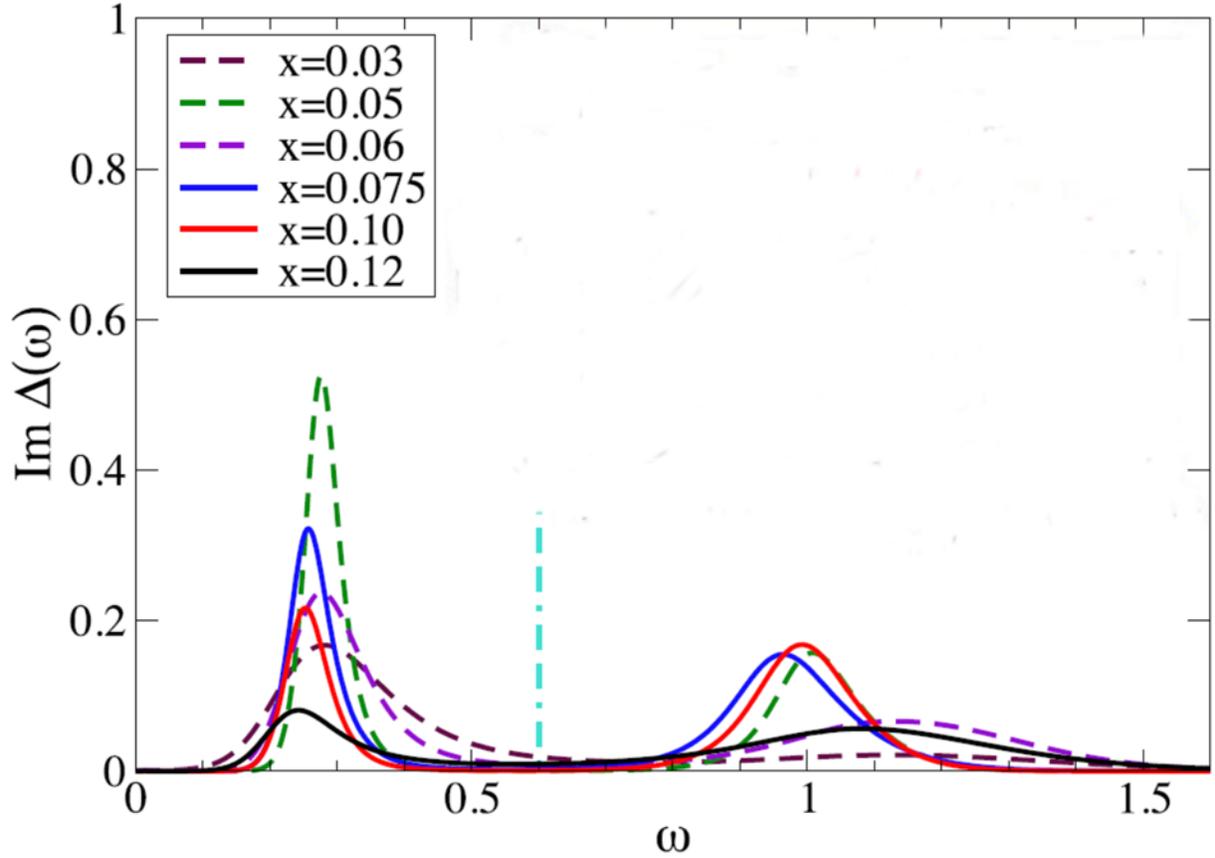

FIG. S4: **Imaginary part of the gap function for the Hubbard model** calculated by Ref. [30] for various dopings indicated in the plot. (The vertical dashed line is used for some other purposes in this reference and does not concern us.) The energy scale $\omega$ is in units of the kinetic energy parameter $t \approx 0.3eV$, chosen to get the maximum $T_c$ as a function of doping to be $t/60$, and the local repulsion parameter $U = 6t$. The gap function $\Delta(\omega)$ and the pairing self-energy $\phi(\omega)$ are related by the quasi-particle renormalization $Z(\omega)$, which is very weakly $\omega$-dependent. The result shown here, specifically the rapid decrease to nearly 0 between the two bumps, does not compare well with the frequency dependence of the experimental pairing self-energy in Fig. 3.